# The Impact of Sodomy Law Repeals on Crime

Riccardo Ciacci[1]

Dario Sansone[2]

**Abstract**

We exploit variation in the timing of decriminalization of same-sex sexual intercourse across U.S. states to estimate the impact of these law changes on crime through difference-in-difference and event-study models. We provide the first evidence that sodomy law repeals led to a decline in the number of arrests for disorderly conduct, prostitution, and other sex offenses. Furthermore, we show that these repeals led to a reduction in arrests for drug and alcohol consumption.

**Keywords:** sodomy laws; LGBTQ; crime

**JEL:** I18; J15; K14; K38

[1] Universidad Pontificia Comillas. E-mail: rciacci@icade.comillas.edu
[2] Vanderbilt University and University of Exeter. E-mail: dario.sansone@vanderbilt.edu

# 1. Introduction

The sexual acts indicated as *sodomy* historically referred to both oral and anal sex, as well as bestiality. Sodomy laws are laws that criminalize these specific sexual activities. American colonies inherited these laws from the British Empire: sodomy was a crime punishable by death in most American colonies. Even after the U.S. declaration of independence and throughout the XX century, sodomy was a crime often punishable by a life sentence. The years after WWI were characterized by a real "gay panic", a widespread belief that homosexuals were sexual predators targeting children and susceptible young adults to make them gay. Between 6,600 and 21,600 people, mostly men, are estimated to have been arrested each year between 1946 and 1961 for non-conforming gender or sexual behaviors. In the same period, tens of thousands of homosexuals were detained, blackmailed, or harassed by police officers (Eskridge 2008). In addition, sodomy laws were used against sexual minorities to limit their rights to adopt or raise children, to justify firing them, and to exclude them from hate-crime laws (ACLU 2019). Before the U.S. Supreme Court deemed sodomy laws unconstitutional in 2003 (*Lawrence v. Texas*), the penalty for violating sodomy laws ranged from a $500 fine in Texas to a maximum life sentence in Idaho (GLAPN 2007).

This paper extends an extremely limited literature on sodomy laws not only in economics, but also in public health and other social sciences. A few studies have looked at the determinants of sodomy laws (Frank, Camp, and Boutcher 2010; Asal, Sommer, and Harwood 2013), or at the effect of legalizing homosexuality across countries on attitudes toward sexual minorities (Kenny and Patel 2017). To our knowledge, there is no study specifically looking at the impact of sodomy laws on crime. The empirical analysis exploits variation in the timing of decriminalization of same-sex sexual intercourse across U.S. states to estimate through difference-in-difference and event-study models. We provide the first evidence that the elimination of sodomy laws led to a persistent decline in the number of arrests for disorderly conduct, prostitution, and other sex offenses. In addition, we show that these repeals led to a reduction in arrests for drug and alcohol consumption.

This paper contributes to two fields. First, within the literature on sexual minorities, this analysis is related to a growing number of studies estimating the impact of LGBT policies such as anti-discrimination laws and same-sex marriage legalization on health and labor market outcomes (Dee 2008; Francis, Mialon, and Peng 2012; Burn 2018; Carpenter et al. 2018; Sansone 2019). Second, this paper is linked to a strand of the literature in crime economics exploring the effect of family and vice laws (Stevenson and Wolfers 2006; Cáceres-Delpiano and Giolito 2012; Heaton 2012). Moreover, inside the field of crime economics, this paper is closely connected to recent empirical studies analyzing sex crimes and/or prostitution (Cunningham and Kendall 2011a; 2011b; Bhuller et al. 2013; Bisschop, Kastoryano, and van der Klaauw 2017; Cunningham and Shah 2018; Ciacci and Sviatschi 2019; Ciacci 2019). Yet, to the best of our knowledge this is the first paper to assess the link these two fields and to estimate the effect of LGBT policies on such crimes.

More generally, this paper provides a new and important contribution to the literature on the economic effects of civil and social right reforms affecting stigmatized and marginalized populations such as the Civil Right Act (J. J. I. Donohue and Heckman 1991; Hersch and Shinall 2015), the legalization of interracial marriage (Fryer 2007), the Americans with Disabilities Act (Acemoglu and Angrist 2001; Hotchkiss 2004), abortion and family-planning reforms (J. J. Donohue and Levitt 2001; Goldin and Katz 2002; Bailey 2006; 2010), and the banning of sex discrimination in schools (Stevenson 2010).

## 2. Institutional context underlying the econometric strategy

Sodomy law decriminalization occurred in two ways: repeal through state legislatures and state supreme court decisions ruling the laws unconstitutional.[3] Before 1980, the call for decriminalization was primarily made by legal experts trying to persuade states to modernize their criminal codes (Eskridge 2008). Illinois became the first state to decriminalize consensual sodomy in 1961. Connecticut did the same in 1969. Slowly, gay and lesbian movement activists, rather than legal experts, became responsible for initiating the attempts to decriminalize sodomy in the last two decades of the 20$^{th}$ century (Bernstein 2003). At the same time, there was also a shift in the primary policy venue used to challenge sodomy laws: as legal activist organizations specializing in judicial challenges began to lead the battle to decriminalize sodomy, they shifted the movement's attention to the judicial system rather than the legislative arena. The move to the courts was largely based on the assumption that judges would be less influenced by public opinion than legislators would, which was particularly important as the federal and state legislatures entered the more conservative Reagan and Bush years (Clendinen and Nagourney, 1999; Kane, 2007). Indeed, historically the U.S. Supreme court had already protect the right to distribute pro-homosexual writing through the public mail service in 1958 (*One Inc. v. Olesen*), while at the same time the federal government was systematically firing during the so-called "Lavender Scare" thousands of U.S. government employees because they were suspected to be homosexual (Johnson 2004).

At the federal court level, the gay and lesbian movement attempted to decriminalize sodomy in the early 1980s through a challenge of the Georgia state sodomy law. The challenge reached the U.S. Supreme Court in 1986 (*Bowers v. Hardwick*). However, by a 5 to 4 decision, the Georgia law was found constitutional and the Court ruled that states had the right to criminalize specific sexual acts. Following this defeat, gay and lesbian activists started to challenge sodomy laws under state constitutions, which can add to rights guaranteed by the U.S. constitution. Thanks to this strategy, homosexuality was decriminalized in Kentucky in 1992 (*Commonwealth v. Wasson*), Tennessee in 1996 (*Campbell v. Sundquist*), and Montana in 1997 (*Gryczan v. Montana*). By the end of 2002, 36 states plus the District of Columbia had decriminalized sodomy in their statutes (GLAPN, 2007; Eskridge, 2008). Finally, the U.S. Supreme Court ruled 6-3 that Texas' sodomy law was unconstitutional (*Lawrence v. Texas*) on June 26, 2003, making all remaining sodomy laws invalid.

---

[3] Table A1 in the Online Appendix provides additional details on the chronology of sodomy laws decriminalization.

## 3. Data and methodology

### 3.1 Data

This paper uses the 1995-2018 Uniform Crime Reporting Program arrest database (FBI 2020).[4] This database collects arrest data for 28 offenses as reported from law enforcement agencies. Since a person might be arrested multiple times in the same year, this dataset measures the number of times persons are arrested rather than the number of individuals arrested.

It is worth noting that the Uniform Crime Reporting Program arrest data set is based on voluntarily reporting by law enforcement agencies. This feature implies there might be the concern that crimes recorded by the database increase simply due to the number of law enforcement agencies that decide to report crimes. To address this issue, we keep track of the number of law enforcement agencies reporting crimes for each state in any year in our sample period.

Table 1 shows summary statistics for our main dependent variables, i.e. arrest rate for disorderly conduct, prostitution, other sex offenses, and driving after consuming alcoholic beverages or using drugs ( per 1,000,000 residents, in logarithms).[5] We can observe that for all four variables mean and median are fairly close to each other. As expected, arrests for prostitution and other sex offenses happen more rarely than arrests for disorderly conduct or for driving under the influence.[6]

### 3.2 Event study model

Given the available data and documented law changes, it is then possible to estimate the following event study:

$$Arrest\_rate_{st} = \alpha + \sum_{k=\underline{T}}^{\overline{T}} \beta_k Sodomy_{st}^k + \delta_s + \mu_t + x'_{st}\gamma_1 + LGBT'_{st}\gamma_2 + \varepsilon_{st}$$

where $arrest_{st}$ is the reported arrest rate (per 1,000,000 residents, in logarithms) for a given crime in state $s$ at time $t$. $Sodomy_{st}^0$ is an indicator equal to one if state $s$ had decriminalized sodomy at time $t$, zero otherwise. $Sodomy_{st}^k$ are the resulting lead ($k < 0$) and lag ($k > 0$) operators. The specification includes state ($\delta_s$) and year ($\mu_t$) fixed effects. The vector of time-varying state-level controls ($x'_{st}$) includes unemployment rate, income per capita, and the number of agencies reporting their crime data to the FBI. In order to control for additional factors potentially related to sodomy laws, $LGBT'_{st}$ accounts for other policies such as constitutional and statutory bans on same-sex marriage, same-sex marriage legalization, same-sex domestic partnership legalization, same-sex civil union legalization, LGBTQ anti-discrimination laws, and LGBTQ hate crime laws. Standard errors are clustered at the state level (Bertrand et al., 2004).

---

[4] We are using all years whose complete arrest reports were available in the FBI UCR website (i.e.,1995-2018): https://ucr.fbi.gov/crime-in-the-u.s (Accessed: August/2020).
[5] All variables used in the empirical analysis are described in detail in Section B of the Online Appendix.
[6] In addition, Table C1 in the Online Appendix displays summary statistics for the number of agencies across states in the considered sample period.

### 3.3 Discussion on the exogeneity of the policy changes

A key concern when interpreting difference-in-difference and event study estimates as causal is that the timing of the sodomy decriminalization in each state should not reflect pre-existing differences in state-level characteristics. In this context, it is important to emphasize that, unlike other policy reforms such as unilateral divorce laws (Stevenson and Wolfers 2006), sodomy laws in the 1990s and early 2000s – i.e., the law changes analyzed in our model - were struck down following judicial decisions, not legislative processes. The exact timing of the court decisions was plausibly unexpected. Moreover, judges often served lengthy terms and were less subject than politicians to the public opinion on homosexuality. Indeed, federal and state judges repealed these sodomy laws at the same time as voters and legislators in several states approved bans on same-sex marriages (Sansone 2019), and while President Clinton and the U.S. Congress passed anti-LGBT legislation such as and the *Defense of Marriage Act* defining marriage for federal purposes as the union of one man and one woman, as well as the *Don't ask, don't tell* policy barring openly gay, lesbian, or bisexual individuals from serving in the military.

It is also worth mentioning that, even if one may worry that the most gay-friendly states were the first ones to introduce LGBTQ reforms such as the legalization of same-sex sexual activity and the introduction of marriage equality, this hypothesis is not supported by the fact that the order in which states decriminalized consensual sodomy is rather different from the order in which states legalized same-sex marriage. For instance, Massachusetts was the first state to legalize same-sex marriage (2004), but it was among the last ones to decriminalize sodomy (2002). New York, one of the states with the largest LGBTQ populations, was not among the first states to legalize sodomy (1980), nor same-sex marriage (2011).

### 4. Results

### 4.1 Sodomy law repeals lead to a reduction in arrest rates

The key finding of the paper is that sodomy law repeals led to a significant and persistent reduction in the arrest rates for crimes directly related to sodomy. Indeed, Figure 1 shows a decline in arrests for sex offenses such as offenses against chastity, common decency, and morals. In line with (Ciacci 2019), Figures 2-3 reports similar reductions in arrests for prostitution and disorderly conduct (i.e., any behavior that tends to disturb the public peace or shock the public sense of morality) respectively. It is worth noting that, in all the graphs, none of the lead operators is statistically significant, thus supporting the parallel trends assumption. Moreover, the impact of decriminalizing sodomy on these crimes can be detected both in the year in which the law was abolished, as well as in the years afterwards, thus suggesting that these reforms had long-term effects.

We then provide evidence supporting the hypothesis that sodomy law decriminalization not only led to a direct decline of individuals arrested for related crimes, but it also had more general effects. In line with the hypothesis that these law changes reduced minority stress (Meyer 1995) and led to a reduction of drinking and drug use as a coping mechanism, Figure 4 reports a clear and significant drop in the number of arrests for driving while mentally or physically impaired as the result of consuming alcoholic beverages or using drugs.

Table 2 reports the results from the difference-in-difference regression model for our four outcome variables (sex offenses, prostitution, disorderly conduct, and driving under the influence). In other words, the table reports the estimated β from the following difference-in-difference model:

$$Arrest\_rate_{st} = \alpha + \beta Sodomy_{st} + \delta_s + \mu_t + x'_{st}\gamma_1 + LGBT'_{st}\gamma_2 + \varepsilon_{st}$$

where $Sodomy_{st}$ is an indicator equal to one if state *s* had decriminalized sodomy at time *t*, as well as in the following years, zero otherwise. The dependent variable ($Arrest\_rate_{st}$), state fixed effects ($\delta_s$), year fixed effects ($\mu_t$), state controls ($x'_{st}$), and LGBT policy controls ($LGBT'_{st}$) are defined as in the event study model. Results are negative and statistically significant in all four regressions, thus supporting the main conclusions from Figures 1-4.

It is worth noting that our estimates are economically meaningful. According to Table 2, our findings suggest that sex offenses and prostitution rates decreased by roughly 16% and 37%, respectively, due to decriminalization of same-sex sexual intercourse. Likewise, disorderly conduct and driving under the influence respectively fell by about 25% each.

### 4.2 Extensions and robustness checks

The Online Appendix also reports several extensions and robustness checks. To further explore the event-study estimates, Columns 1 of Tables C2-C5 in the Online Appendix show the estimated coefficients shown in Figures 1-4. As also evident from these figures, most of the estimated effects or arrest rates in the years sodomy laws were repealed and afterwards are statistically significant at the 5 percent or 1 percent levels. In addition, following Borusyak and Jaravel (2018), Columns 2-3 of such tables omit either $Sodomy_{st}^{-2}$ or $Sodomy_{st}^{-3}$, respectively. The last row of these two columns displays the p-value of the F-test of significance of the remaining lead operator: we find no statistical evidence suggesting that our results are driven by pre-trends.

The main results do not change when measuring arrests in levels rather than logarithms (Figures C1-C4), when restricting the time frame (Figure C5-C8), or when increasing the number of leads and lags (Figure C9-C12). Excluding California – the state with the largest number of LGBT individuals – does not substantially alter the main findings (Figures C13-16). In line with the estimates plotted in Figure 4 measuring the impact of sodomy law repeals on arrest rate for driving after consuming alcoholic beverages or using drugs, we observe similar reductions in the number of arrests for drug abuse (Figure C17) and liquor laws violations (Figure C18). Finally, we show as placebo tests that sodomy law repeals had no impact on the number of arrests for gambling (Figure C19), burglary (Figure C20), or arson (Figure C21).

### 5. Conclusions

This paper has provided the first evidence that sodomy law repeals had an economic impact: they led to a reduction in the number of arrests due to sex offenses, prostitution, or disorderly conduct, as well as a decline in arrests linked to alcohol and drug consumption.

These findings are important from a policy perspective. Institutionalized homophobia is still prevalent worldwide: as of 2020, 70 countries have laws criminalizing homosexuality. In 11 of these countries, homosexuality is punishable by death (ILGA 2019). This study is a first step towards helping international institutions such as the World Bank or the European Union evaluate more accurately the costs and benefits of suspending foreign aids to countries in blatant violation of basic human rights (Economist 2014; Steer 2018). Furthermore, this analysis emphasizes the potential benefits from repealing sodomy laws still standing in several countries.

**Figure 1: Effect of sodomy law repeals on arrests for sex offenses.**

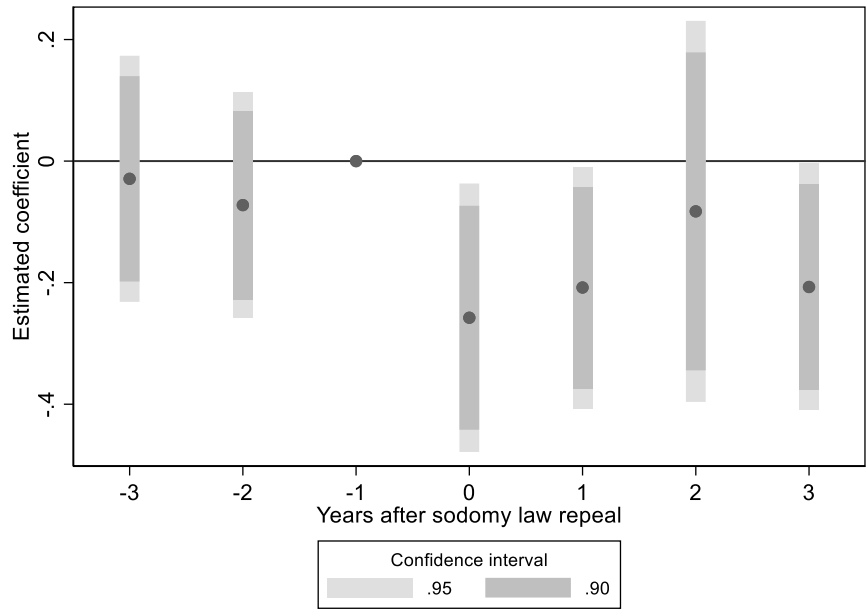

This figure analyzes the effect of sodomy law repeals on the arrest rate for sex offenses (excluding rape, prostitution, and commercial vice). Arrest rate (per 1,000,000 state residents) is in logarithms. First lead normalized to zero. See also Data and Methodology Section. Source: FBI 1995-2018. N=1,189.

**Figure 2: Effect of sodomy law repeals on arrests for prostitution.**

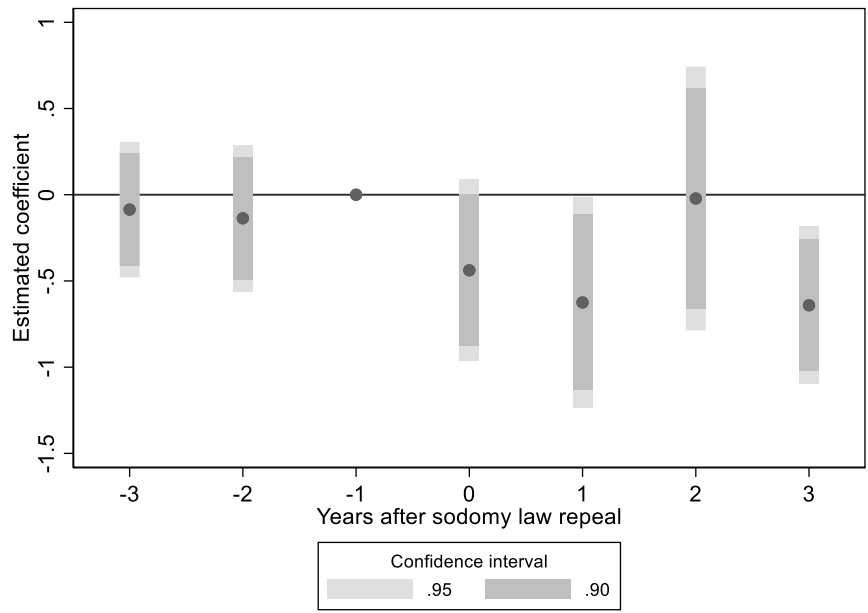

This figure analyzes the effect of sodomy law repeals on the arrest rate (in logarithm) for prostitution and commercialized vice. See notes in Figure 1. Source: FBI 1995-2018. N=1,188.

**Figure 3: Effect of sodomy law repeals on arrests for disorderly conduct.**

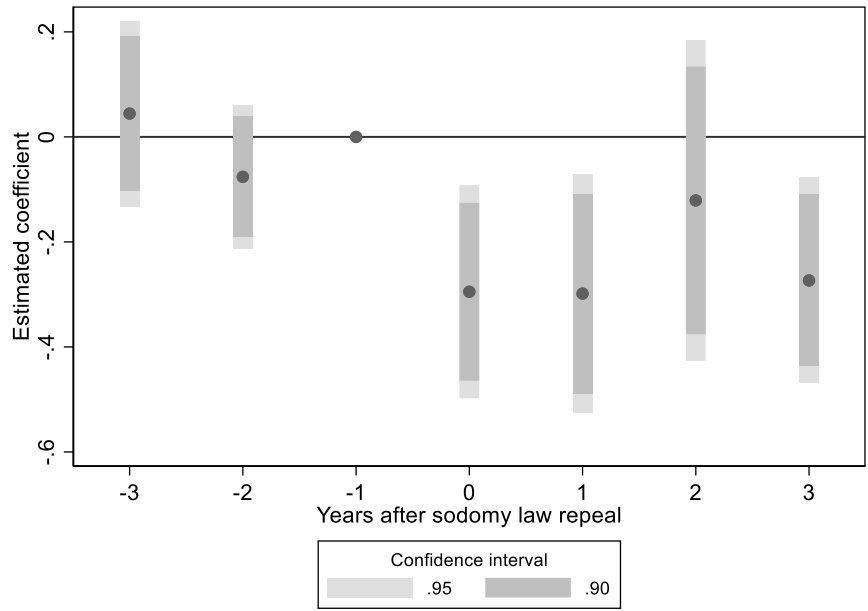

This figure analyzes the effect of sodomy law repeals on the arrest rate (in logarithm) for disorderly conduct. See notes in Figure 1. Source: FBI 1995-2018. N=1,179.

**Figure 4: Effect of sodomy law repeals on arrests for driving under the influence.**

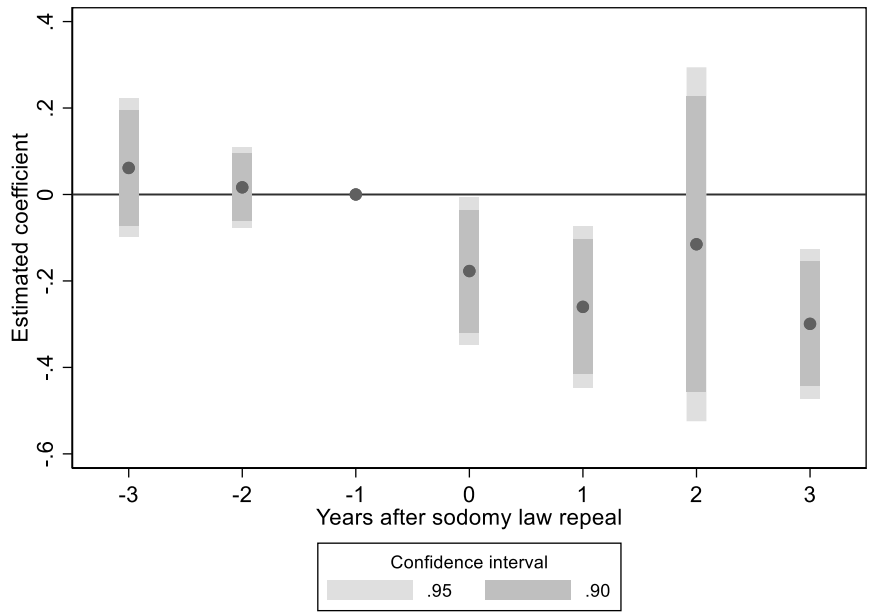

This figure analyzes the effect of sodomy law repeals on the arrest rate (in logarithm) for driving after consuming alcoholic beverages or using drugs. See notes in Figure 1. Source: FBI 1995-2018. N=1,188.

**Table 1: Descriptive statistics**.

|                              | Mean | St. Dev. | Min | Max  | Median |
|------------------------------|------|----------|-----|------|--------|
| Sex offenses                 | 5.23 | 0.91     | 0   | 6.91 | 5.33   |
| Prostitution                 | 4.58 | 1.50     | 0   | 8.01 | 4.86   |
| Disorderly conduct           | 7.40 | 1.09     | 0   | 9.82 | 7.54   |
| Driving under the influence  | 8.26 | 0.96     | 0   | 9.77 | 8.43   |

This table displays descriptive statistics for the main dependent variables during the considered sample period. Source: FBI 1995-2018.

**Table 2: Effect of sodomy law repeals on sex offenses, prostitution, and disorderly conduct. Difference-in-difference.**

|                       | Sex offenses | Prostitution | Disorderly conduct | Driving under the influence |
|-----------------------|--------------|--------------|--------------------|------------------------------|
|                       | (1)          | (2)          | (3)                | (4)                          |
| Sodomy law repeal     | -0.170*      | -0.464**     | -0.277***          | -0.297***                    |
|                       | (0.094)      | (0.176)      | (0.092)            | (0.079)                      |
| State fixed effects   | ✓            | ✓            | ✓                  | ✓                            |
| Year fixed effects    | ✓            | ✓            | ✓                  | ✓                            |
| State control         | ✓            | ✓            | ✓                  | ✓                            |
| LGBT policies         | ✓            | ✓            | ✓                  | ✓                            |
| Observations          | 1,189        | 1,188        | 1,179              | 1,188                        |
| Adjusted-$R^2$        | 0.762        | 0.681        | 0.822              | 0.805                        |

This table analyzes the effect of sodomy law repeals on the arrest rate for sex offenses other than rape or prostitution (Column 1); prostitution and commercialized vice (Column 2); disorderly conduct (Column 3); and driving under the influence of alcohol or drugs (Column 4). Arrest rate (per 1,000,000 state residents) is in logarithms. Time-varying state-level controls: unemployment rate, income per capita, and the number of agencies reporting their crime data to the FBI. LGBT policies: constitutional and statutory bans on same-sex marriage, same-sex marriage legalization, same-sex domestic partnership legalization, same-sex civil union legalization, LGBTQ anti-discrimination laws, and LGBTQ hate crime laws. Standard errors clustered at the state level reported in parenthesis. Source: FBI 1995-2018. * $p < 0.10$, ** $p < 0.05$, *** $p < 0.01$

# The Impact of Sodomy Law Repeals on Crime

Riccardo Ciacci[7]

Dario Sansone[8]

**Online Appendix**

---

[7] Universidad Pontificia Comillas. E-mail: rciacci@icade.comillas.edu
[8] Vanderbilt University and University of Exeter. E-mail: dario.sansone@vanderbilt.edu

# Appendix A. Institutional context underlying the econometric strategy

**Table A1: Sodomy law repeal before *Lawrence v. Texas* (2003).**

| State | Year | Method | Notes |
|---|---|---|---|
| Illinois | 1961 | Legislative | Enacted in 1961, effective in 1962 |
| Connecticut | 1969 | Legislative | Enacted in 1969, effective in 1971 |
| Colorado | 1971 | Legislative | Enacted in 1971, effective in 1972 |
| Oregon | 1971 | Legislative | Enacted in 1971, effective in 1972 |
| Delaware | 1972 | Legislative | Enacted in 1972, effective in 1973 |
| Hawaii | 1972 | Legislative | Enacted in 1972, effective in 1973 |
| Ohio | 1972 | Legislative | Enacted in 1972, effective in 1974 |
| North Dakota | 1973 | Legislative | Enacted in 1973, effective in 1975 |
| California | 1975 | Legislative | Enacted in 1975, effective in 1976 |
| Maine | 1975 | Legislative | Enacted in 1975, effective in 1976 |
| New Hampshire | 1975 | Legislative | Enacted in 1975, effective in 1975 |
| New Mexico | 1975 | Legislative | Enacted in 1975, effective in 1975 |
| Washington | 1975 | Legislative | Enacted in 1975, effective in 1976 |
| Indiana | 1976 | Legislative | Enacted in 1976, effective in 1977 |
| Iowa | 1976 | Legislative | Enacted in 1976, effective in 1978 |
| South Dakota | 1976 | Legislative | Enacted in 1976, effective in 1977 |
| West Virginia | 1976 | Legislative | Enacted in 1976, effective in 1976 |
| Nebraska | 1977 | Legislative | Enacted in 1977, effective in 1978 |
| Vermont | 1977 | Legislative | Enacted in 1977, effective in 1977 |
| Wyoming | 1977 | Legislative | Enacted in 1977, effective in 1977 |
| Alaska | 1978 | Legislative | Enacted in 1978, effective in 1980 |
| New Jersey | 1978 | Legislative | Enacted in 1978, effective in 1979 |
| New York | 1980 | Judicial | *New York v. Onofre* |
| Pennsylvania | 1980 | Judicial | *Commonwealth v. Bonadio* |
| Wisconsin | 1983 | Legislative | Enacted in 1983, effective in 1983 |
| Kentucky | 1992 | Judicial | *Commonwealth v. Wasson* |
| DC | 1993 | Legislative | Enacted in 1993, effective in 1994 |
| Nevada | 1993 | Legislative | Enacted in 1993, effective in 1993 |
| Tennessee | 1996 | Judicial | *Campbell v. Sundquist* |
| Montana | 1997 | Montana | *Gryczan v. Montana* |
| Georgia | 1998 | Judicial | *Powell v. Georgia* |
| Rhode Island | 1998 | Legislative | Enacted in 1998, effective in 1998 |
| Maryland | 1999 | Judicial | *Williams v. Glendening* |
| Arizona | 2001 | Legislative | Enacted in 2001, effective in 2001 |
| Minnesota | 2001 | Judicial | *Doe et al. v. Ventura et al.* |
| Arkansas | 2002 | Judicial | *Jegley v. Picado* |
| Massachusetts | 2002 | Judicial | *GLAD v. Attorney General* |

Main Source: GLAPN (2007); Kane (2007); Eskridge (2008).

**Appendix B. Variable description.**

**B.1 Key variables.**

*Number of arrests.* The Uniform Crime Report (UCR) Program Data is a collection of agency-level data published by the FBI. The FBI website reports complete UCR annual data for the years 1995-2018.[9] Because a person may be arrested multiple times during a year, the UCR arrest figures do not reflect the number of individuals who have been arrested; rather, the arrest data show the number of times that persons are arrested, as reported by law enforcement agencies to the UCR Program. We have analyzed the following crimes by dividing the number of reported arrests by the state population:

- Prostitution and commercialized vice: unlawful promotion of or participation in sexual activities for profit.
- Sex offenses (except rape, prostitution, and commercialized vice): Offenses against chastity, common decency, morals, and the like.
- Disorderly conduct: any behavior that tends to disturb the public peace or decorum, scandalize the community, or shock the public sense of morality.
- Driving under the influence: driving or operating a motor vehicle or common carrier while mentally or physically impaired as the result of consuming an alcoholic beverage or using a drug or narcotic.
- Liquor laws: the violation of state or local laws or ordinances prohibiting the manufacture, sale, purchase, transportation, possession, or use of alcoholic beverages, not including driving under the influence and drunkenness. Federal violations are excluded.
- Drug abuse violations: violation of laws prohibiting the production, distribution, and/or use of certain controlled substances. This includes the unlawful cultivation, manufacture, distribution, sale, purchase, use, possession, transportation, or importation of any controlled drug or narcotic substance. The following drug categories are specified: opium or cocaine and their derivatives (morphine, heroin, codeine); marijuana; synthetic narcotics, i.e. manufactured narcotics that can cause true addiction (Demerol, methadone); and dangerous nonnarcotic drugs (barbiturates, Benzedrine).
- Burglary: the unlawful entry of a structure to commit a felony or theft. To classify an offense as a burglary, the use of force to gain entry need not have occurred.
- Gambling: to unlawfully bet or wager money or something else of value; assist, promote, or operate a game of chance for money or some other stake; possess or transmit wagering information; manufacture, sell, purchase, possess, or transport gambling equipment, devices, or goods; or tamper with the outcome of a sporting event or contest to gain a gambling advantage.

---

[9] Source: https://ucr.fbi.gov/crime-in-the-u.s/. Accessed: Mar/1/2020

- Arson: any willful or malicious burning or attempting to burn, with or without intent to defraud, a dwelling house, public building, motor vehicle or aircraft, personal property of another, etc.

*Population* records the estimates of the civilian noninstitutional population ages 16 and older computed by the Census Bureau.[10]

*Sodomy law repeal* is an indicator variable equal to one in all states and time periods in which sodomy laws regarding same-sex sexual activities (both oral and anal sex) had been repealed\decriminalized; zero otherwise. This variable has been set equal to one even in cases when a state or federal Supreme Court had found sodomy laws unconstitutional, although sodomy laws were still included in the state statute, since they were inapplicable. The enactment date has been used to code this variable (as shown in Table A1, all sodomy laws repealed in the time frame considered in the main analysis, i.e. 1995-2018, have the effective date in the same years as the enactment date). Whenever noted, some minor variations of this variables have been used in the event studies and difference-in-difference models. These data have been primarily obtained from the Gay and Lesbian Archives of the Pacific Northwest.[11]

**B.2 State-level controls.**

*Number of agencies* records in each year and state the number of agencies that reported their crime statistics to the UCR.

*Unemployment rate* records the state-month unemployment rates for the civilian noninstitutional population ages 16 and older, not seasonally adjusted as computed from the Bureau of Labor Statistics.[12] From this, we have computed the average unemployment rate in each state.

*Income per capita* records the state-year personal income, not seasonally adjusted. The data have been retrieved from FRED, Federal Reserve Bank of St. Louis.[13]

**B.3 LGBT policy variables.**

*SSM legal* is an indicator variable equal to one in all states and time periods when same-sex marriage was legal; zero otherwise. The effective date has been used to code this variable. These data have been primarily obtained from the National Center for Lesbian Rights.[14]

*SSM ban* is a series of indicator variables equal to one in all states and time periods in which same-sex marriage was banned in the state constitution or state statute; zero otherwise. These indicators remain equal to one even in later years after the legalization of same-sex marriage in a given state. When more than one statutory ban was passed in a state, the oldest one has been used to code the

---

[10] Source: https://www.bls.gov/lau/rdscnp16.htm. Accessed: Oct/1/2019.
[11] Source: https://www.glapn.org/sodomylaws/usa/usa.htm. Accessed Oct/1/2019.
[12] Source: https://www.bls.gov/lau/rdscnp16.htm. Accessed: Oct/1/2019.
[13] Applied filters: income; not seasonally adjusted, per capita, state. Source: https://fred.stlouisfed.org/. Accessed: Oct/25/2019
[14] Source: http://www.nclrights.org/wp-content/uploads/2015/07/Relationship-Recognition.pdf. Accessed Oct/1/2019.

state statute ban variable. These data have been primarily obtained from the Freedom to Marry campaign.[15]

*Domestic partnership* is an indicator variable equal to one in all states and time periods in which same-sex domestic partnerships were legal; zero otherwise. This indicator remains equal to one even in later years when\if a state had converted same-sex domestic partnerships into marriages. These data have been primarily obtained from the National Center for Lesbian Rights.[16]

*Civil union* is an indicator variable equal to one in all states and time periods in which same-sex civil unions were legal; zero otherwise. This indicator remains equal to one even in later years when\if a state had converted same-sex civil unions in marriages. These data have been primarily obtained from the National Center for Lesbian Rights.[17]

*Anti-discrimination law* is an indicator equal to one in all states and time periods in which employer discrimination based on sexual orientation was not allowed; zero otherwise. This variable has been set equal to one even if the law covered only sexual orientation, not gender identity, or if a law protecting trans individuals was passed at a later date. Laws protecting only public employees have not been considered. These data have been primarily obtained from the Freedom for All Americans campaign.[18]

*Hate crime* is a series of indicator variables equal to one in all states and time periods in which there was a law specifically addressing hate or bias crimes based on sexual orientation only, or on sexual orientation and gender identity; zero otherwise. Since some states passed these laws after 2009, these variables have not been set equal to one for all states after President Obama signed the Matthew Shepard and James Byrd, Jr. Hate Crimes Prevention Act into law on October 28, 2009. These data have been primarily obtained from the Human Rights Campaign.[19]

---

[15] Source: http://www.freedomtomarry.org/pages/winning-in-the-states. Accessed Oct/1/2019.
[16] Source: http://www.nclrights.org/wp-content/uploads/2015/07/Relationship-Recognition.pdf. Accessed Oct/1/2019.
[17] Source: http://www.nclrights.org/wp-content/uploads/2015/07/Relationship-Recognition.pdf. Accessed Oct/1/2019.
[18] Source: https://www.freedomforallamericans.org/states/.Accessed: Oct/21/2019.
[19] Source: https://www.hrc.org/state-maps/hate-crimes. Accessed: Oct/25/2019.

# Appendix C. Additional tables and figures.

**Figure C1: Effect of sodomy law repeals on arrests for sex offenses (in levels).**

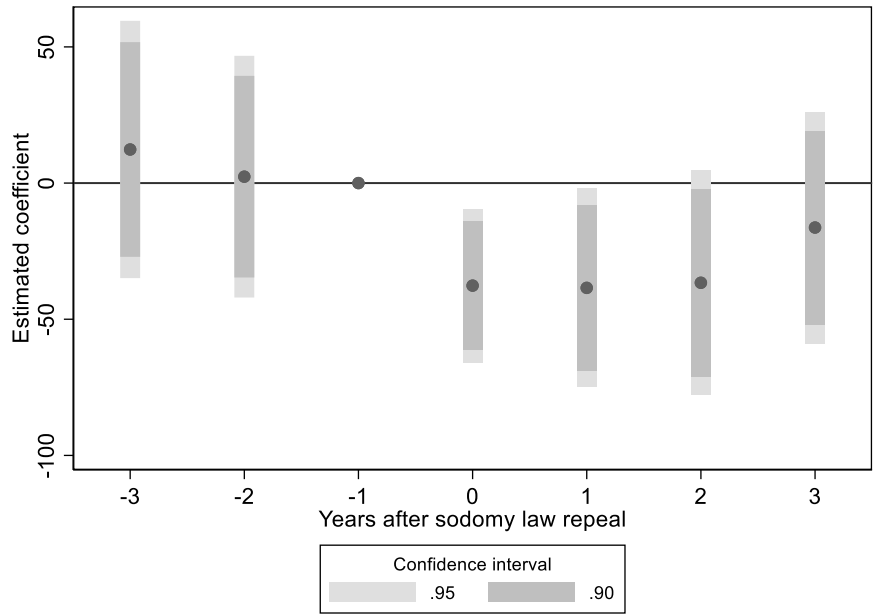

This figure analyzes the effect of sodomy law repeals on the arrest rate (in levels) for sex offenses (excluding rape, prostitution, and commercial vice). See also notes in Figure 1. Source: FBI 1995-2018. N=1,189.

**Figure C2: Effect of sodomy law repeals on arrests for prostitution (in levels).**

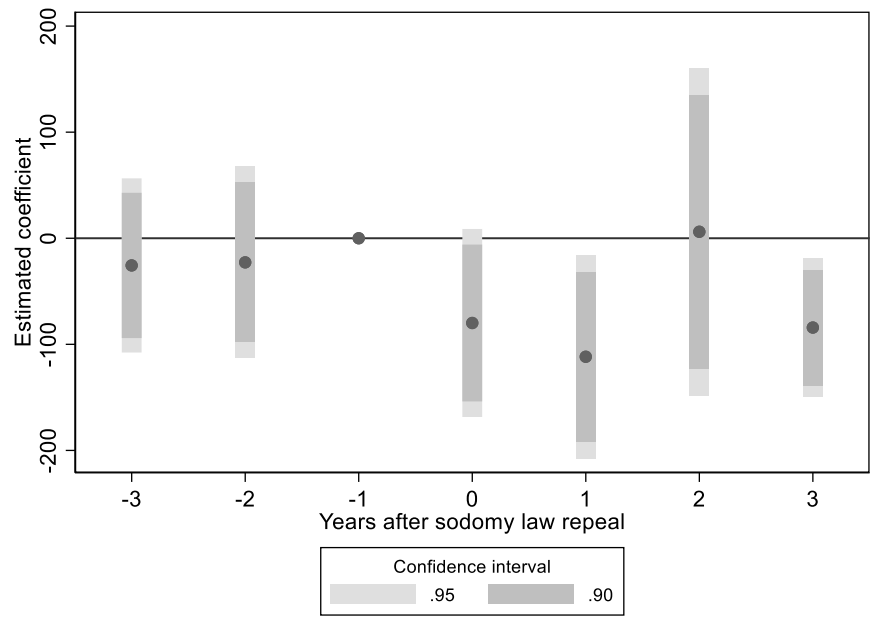

This figure analyzes the effect of sodomy law repeals on the arrest rate (in levels) for prostitution and commercialized vice. See also notes in Figure 1. Source: FBI 1995-2018. N=1,188.

**Figure C3: Effect of sodomy law repeals on arrests for disorderly conduct (in levels).**

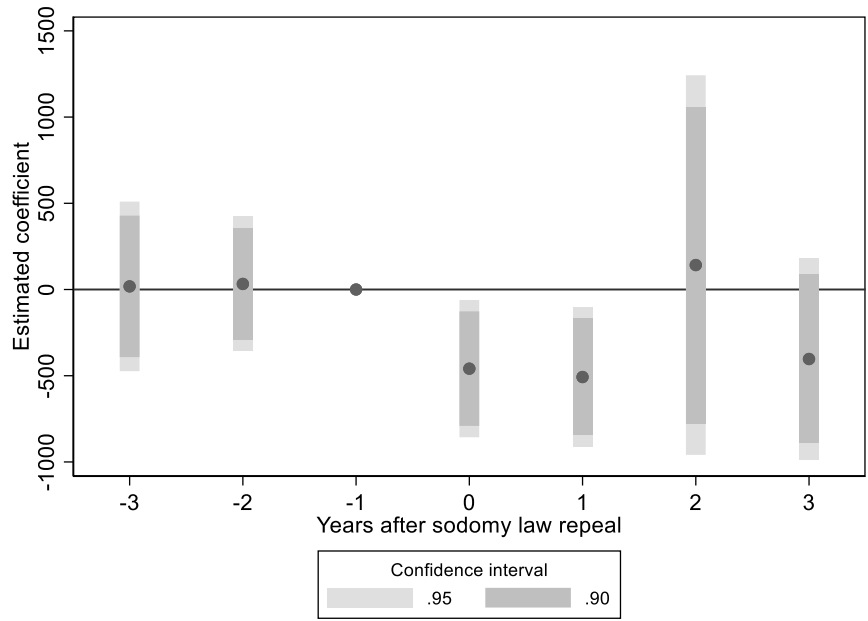

This figure analyzes the effect of sodomy law repeals on the arrest rate (in levels) for disorderly conduct. See also notes in Figure 1. Source: FBI 1995-2018. N=1,179.

**Figure C4: Effect of sodomy law repeals on arrests for driving under the influence (in levels).**

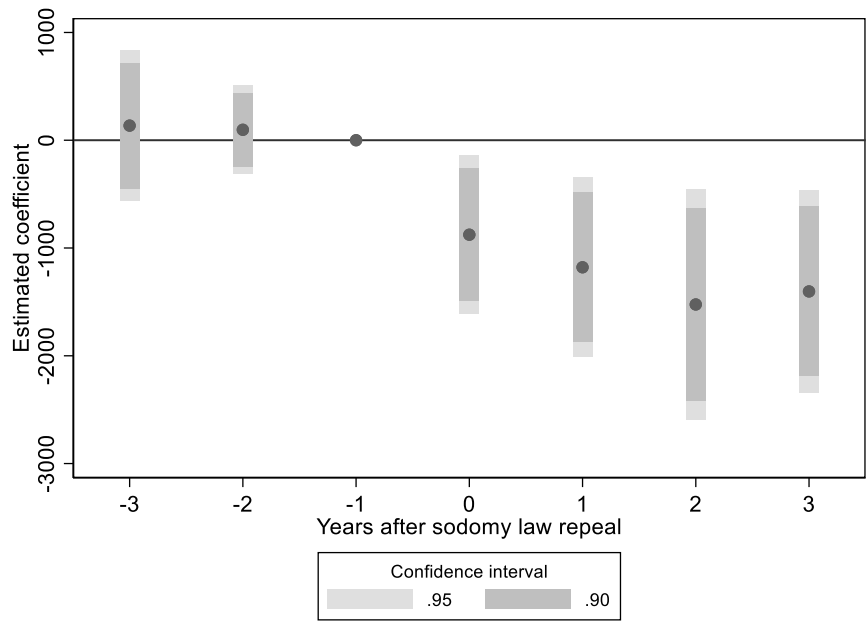

This figure analyzes the effect of sodomy law repeals on the arrest rate (in levels) for driving after consuming alcoholic beverages or using drugs. See notes in Figure 1. Source: FBI 1995-2018. N=1,188.

**Figure C5: Effect of sodomy law repeals on arrests for sex offenses (1995-2010).**

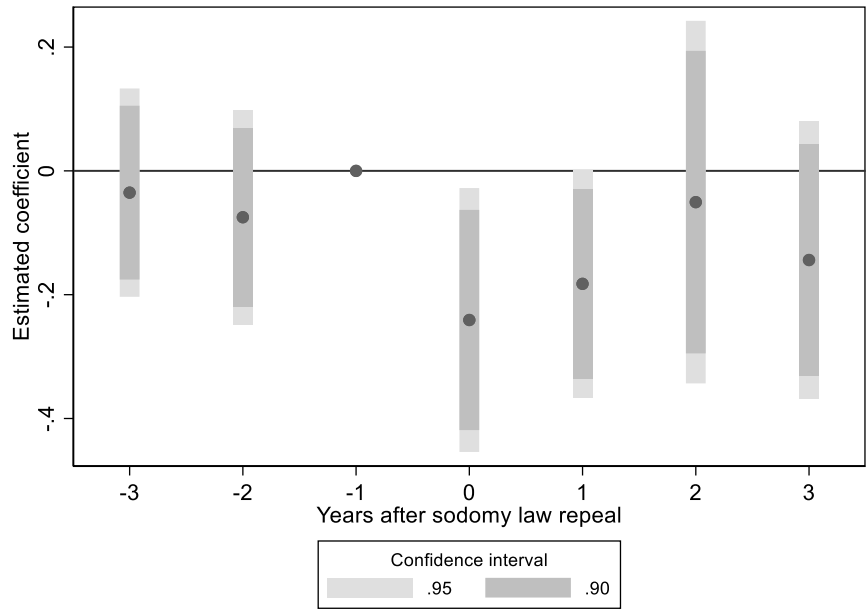

This figure analyzes the effect of sodomy law repeals on arrest rate (in logarithm) for sex offenses (excluding rape, prostitution, and commercial vice). See also notes in Figure 1. Source: FBI 1995-2010. N=784.

**Figure C6: Effect of sodomy law repeals on arrests for prostitution (1995-2010).**

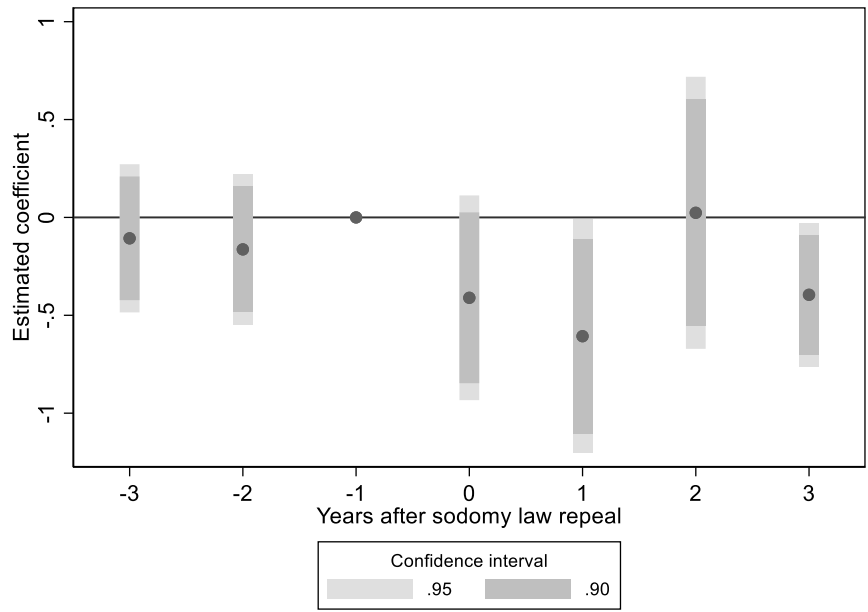

This figure analyzes the effect of sodomy law repeals on the arrest rate (in logarithm) for prostitution and commercialized vice. See also notes in Figure 1. Source: FBI 1995-2010. N=783.

**Figure C7: Effect of sodomy law repeals on arrests for disorderly conduct (1995-2010).**

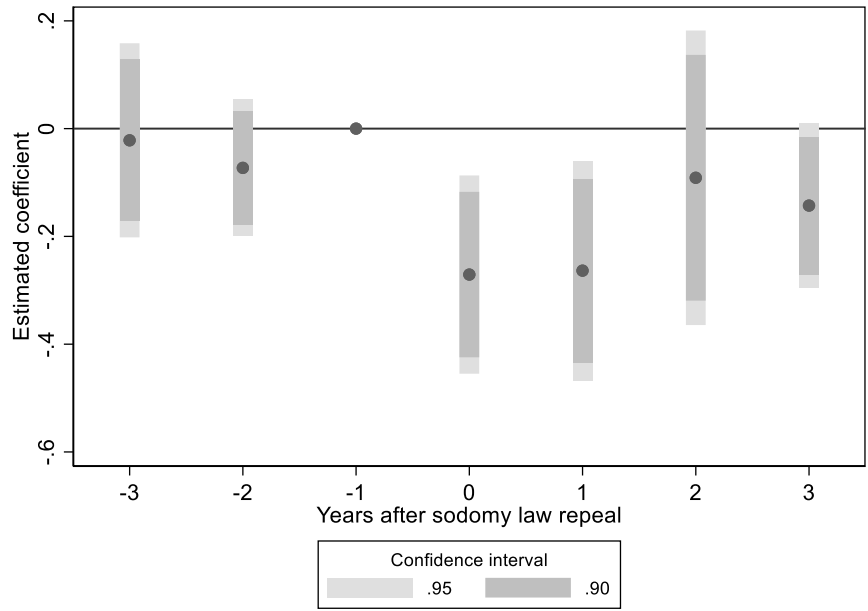

This figure analyzes the effect of sodomy law repeals on the arrest rate (in logarithm) for disorderly conduct. See also notes in Figure 1. Source: FBI 1995-2010. N=774.

**Figure C8: Effect of sodomy law repeals on arrests for driving under the influence (1995-2010).**

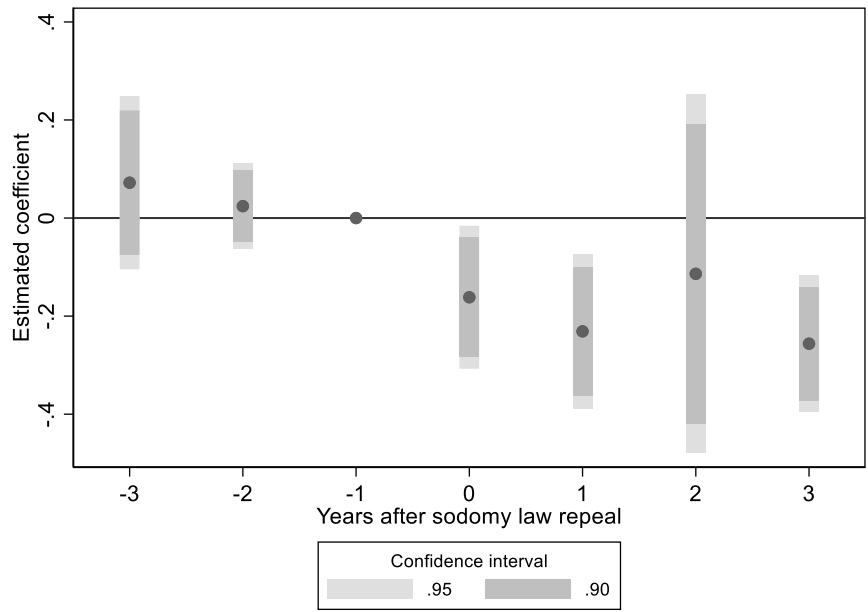

This figure analyzes the effect of sodomy law repeals on the arrest rate (in logarithm) for driving after consuming alcoholic beverages or using drugs. See notes in Figure 1. Source: FBI 1995-2010. N=783.

**Figure C9: Effect of sodomy law repeal on arrests for sex offenses. Add leads and lags.**

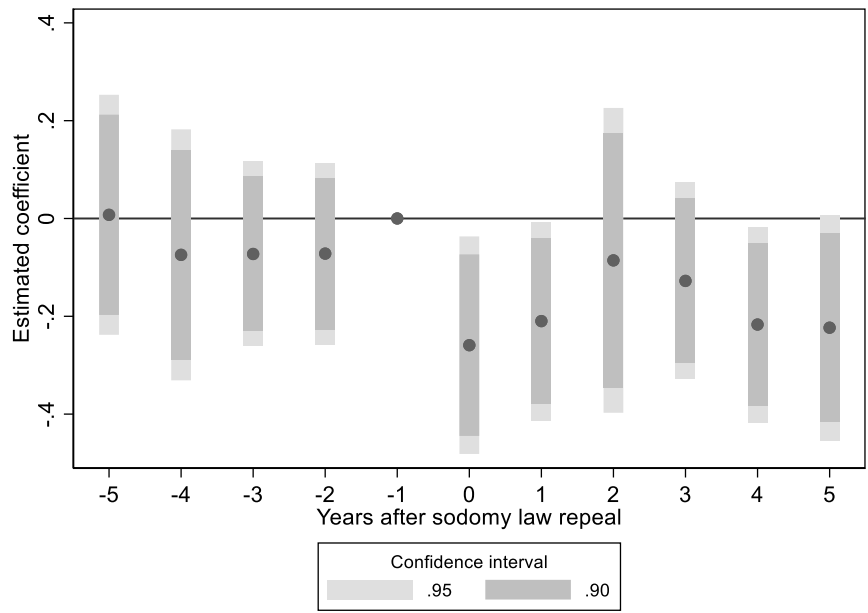

This figure analyzes the effect of sodomy law repeals on the arrest rate (in logarithm) for sex offenses (excluding rape, prostitution, and commercial vice). See also notes in Figure 1. Source: FBI 1995-2018. N=1,189.

**Figure C10: Effect of sodomy law repeals on arrests for prostitution. Add leads and lags.**

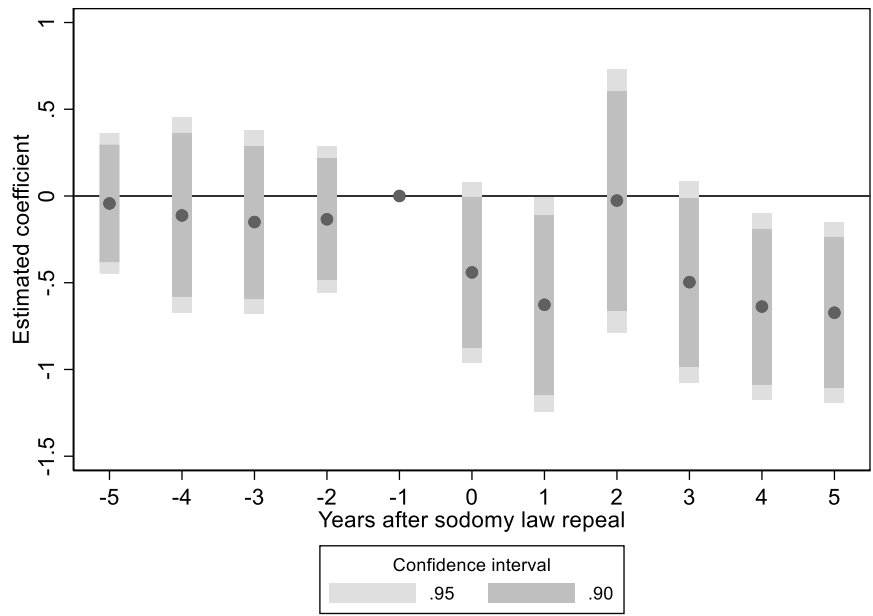

This figure analyzes the effect of sodomy law repeals on the arrest rate (in logarithm) for prostitution and commercialized vice. See also notes in Figure 1. Source: FBI 1995-2018. Source: FBI 1995-2018. N=1,188.

**Figure C11: Effect of sodomy law repeals on arrests for disorderly conduct. Add leads and lags.**

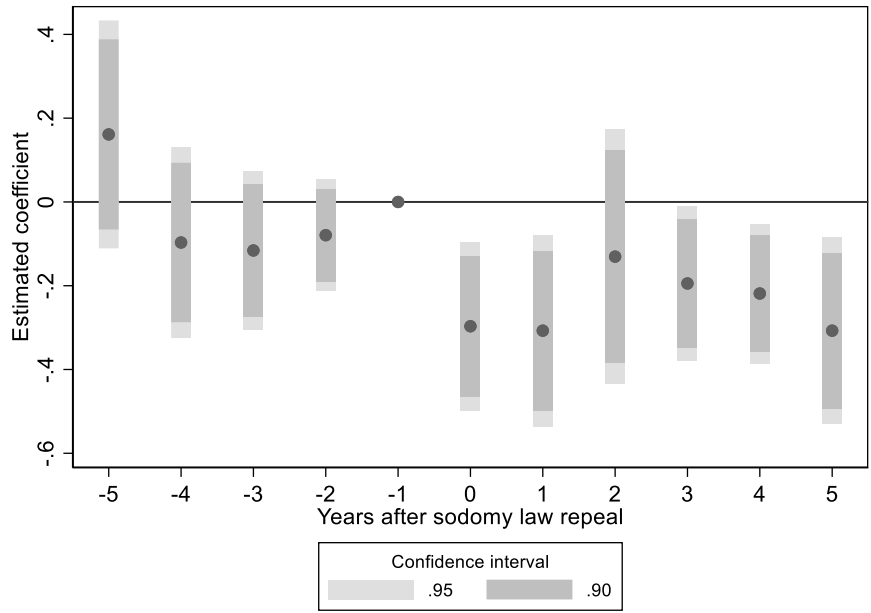

This figure analyzes the effect of sodomy law repeals on the arrest rate (in logarithm) for disorderly conduct. See also notes in Figure 1. Source: FBI 1995-2018. N=1,179.

**Figure C12: Effect of sodomy law repeals on arrests for driving under the influence. Add leads and lags.**

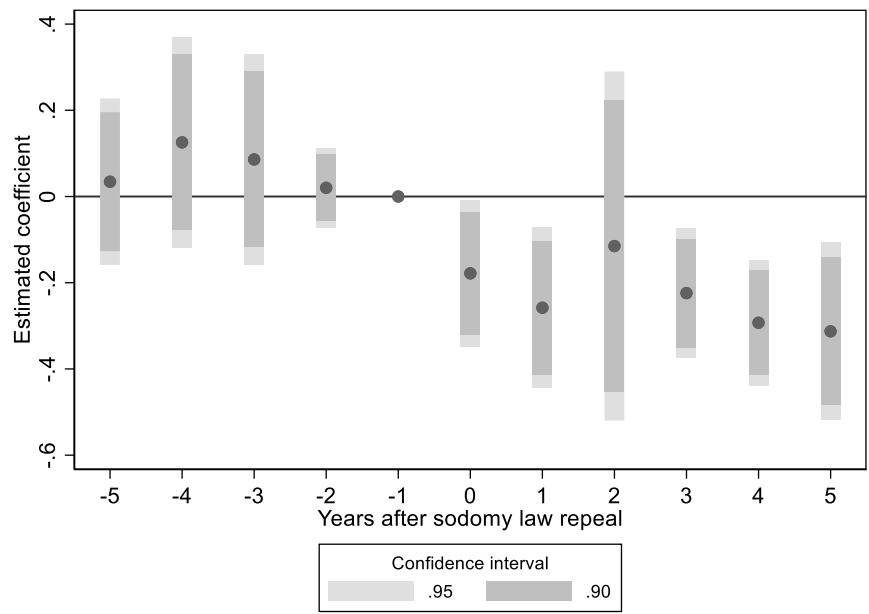

This figure analyzes the effect of sodomy law repeals on the arrest rate (in logarithm) for driving after consuming alcoholic beverages or using drugs. See notes in Figure 1. Source: FBI 1995-2018. N=1,188.

**Figure C13: Effect of sodomy law repeal on arrests for sex offenses. Exclude California.**

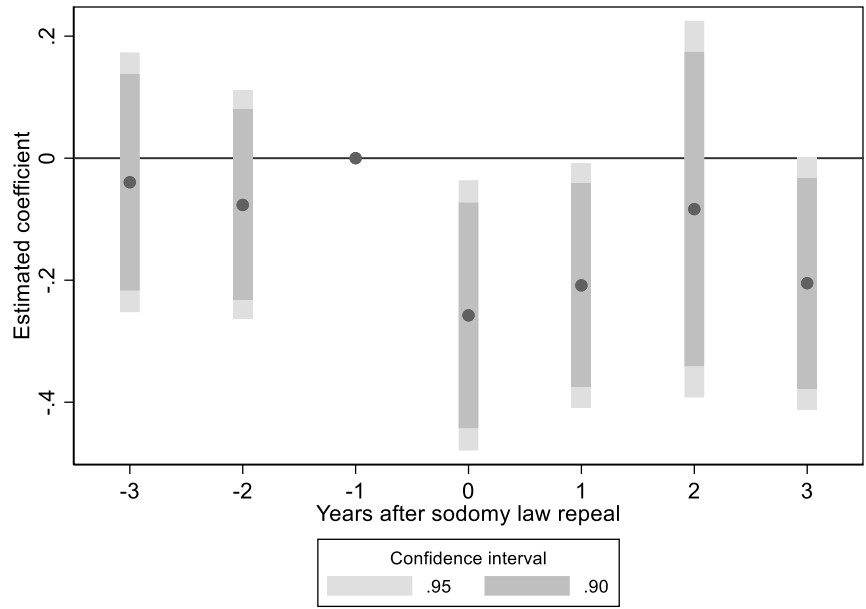

This figure analyzes the effect of sodomy law repeals on the arrest rate (in logarithm) for sex offenses (excluding rape, prostitution, and commercial vice). See also notes in Figure 1. Source: FBI 1995-2018. N=1,165.

**Figure C14: Effect of sodomy law repeals on arrests for prostitution. Exclude California.**

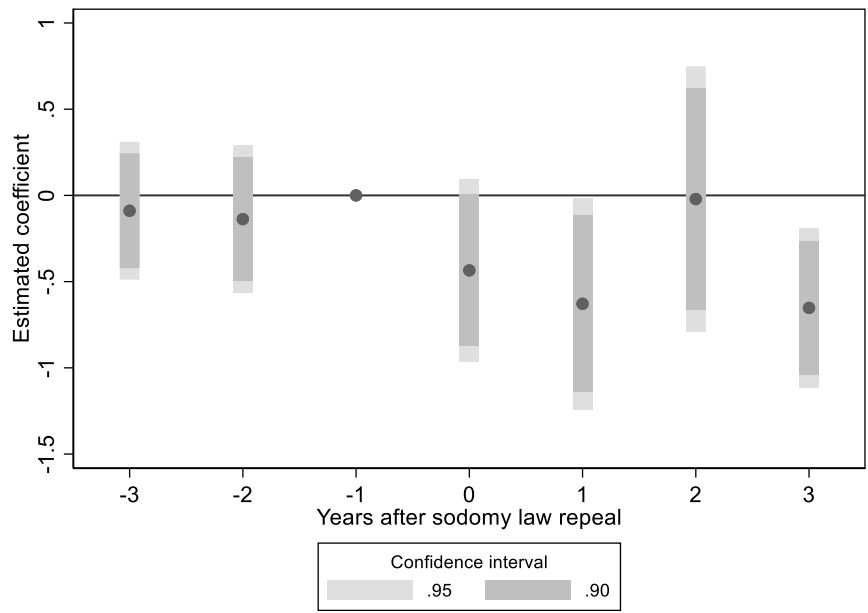

This figure analyzes the effect of sodomy law repeals on the arrest rate (in logarithm) for prostitution and commercialized vice. See also notes in Figure 1. Source: FBI 1995-2018. Source: FBI 1995-2018. N=1,164.

**Figure C15: Effect of sodomy law repeals on arrests for disorderly conduct. Exclude California.**

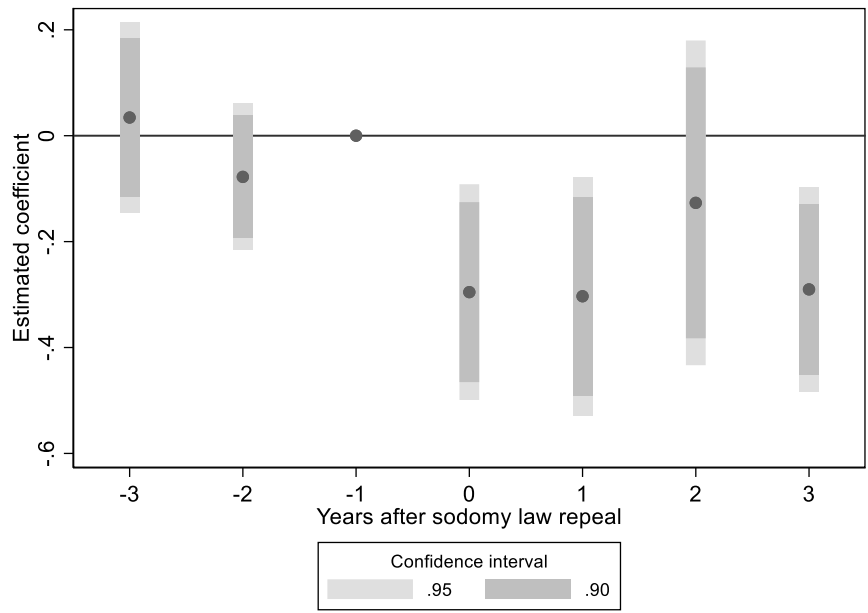

This figure analyzes the effect of sodomy law repeals on the arrest rate (in logarithm) for disorderly conduct. See also notes in Figure 1. Source: FBI 1995-2018. N=1,155.

**Figure C16: Effect of sodomy law repeals on arrests for driving under the influence. Exclude California.**

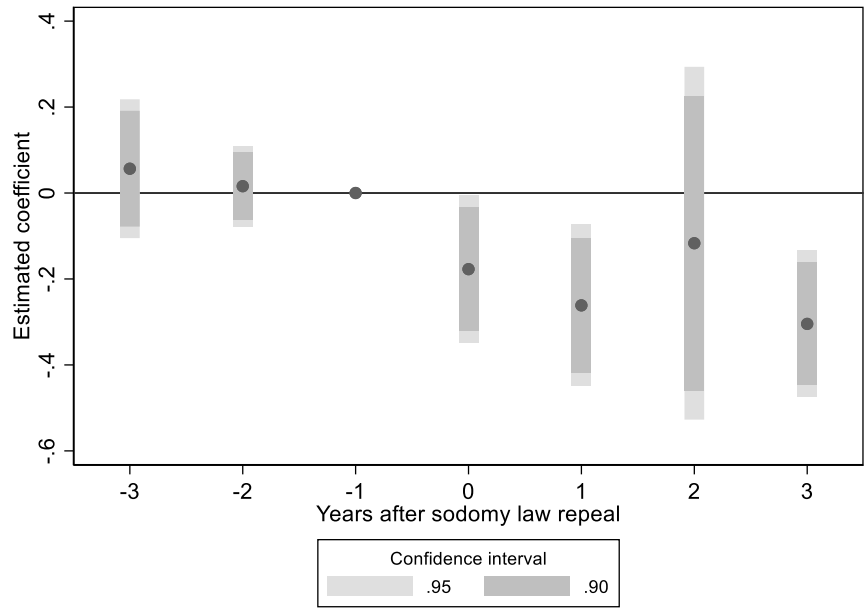

This figure analyzes the effect of sodomy law repeals on the arrest rate (in logarithm) for driving after consuming alcoholic beverages or using drugs. See notes in Figure 1. Source: FBI 1995-2018. N=1,164.

**Figure C17: Effect of sodomy law repeals on arrests for drug abuse.**

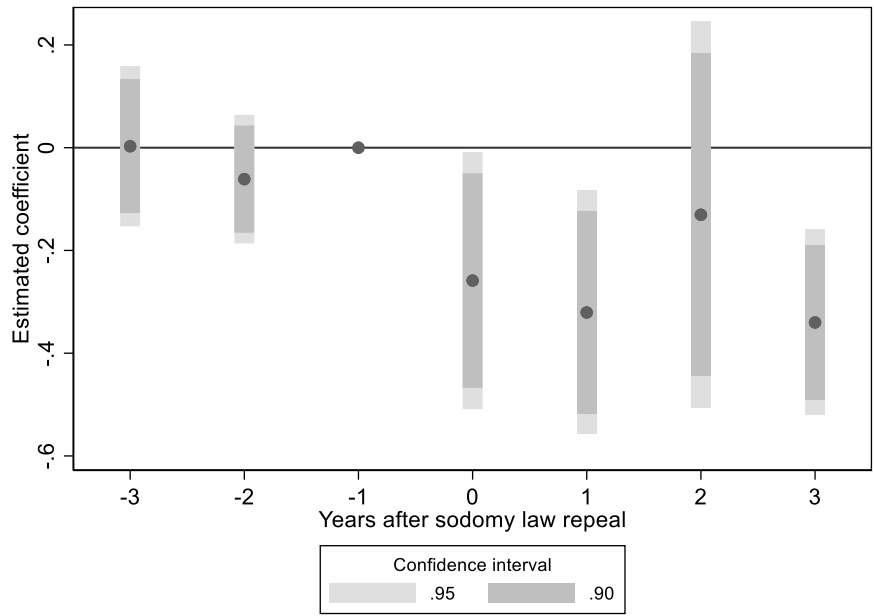

This figure analyzes the effect of sodomy law repeals on the arrest rate (in logarithm) for drug abuse violations. See also notes in Figure 1. Source: FBI 1995-2018. N=1,189.

**Figure C18: Effect of sodomy law repeals on arrests for liquor law violations.**

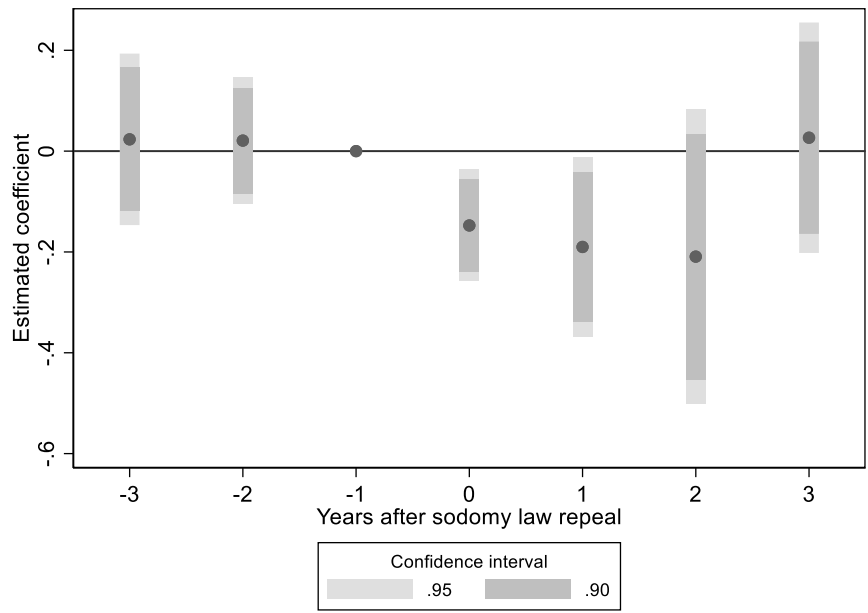

This figure analyzes the effect of sodomy law repeals on the arrest rate (in logarithm) for liquor law violations. See also notes in Figure 1. Source: FBI 1995-2018. N=1,189.

**Figure C19: Effect of sodomy law repeals on arrests for gambling.**

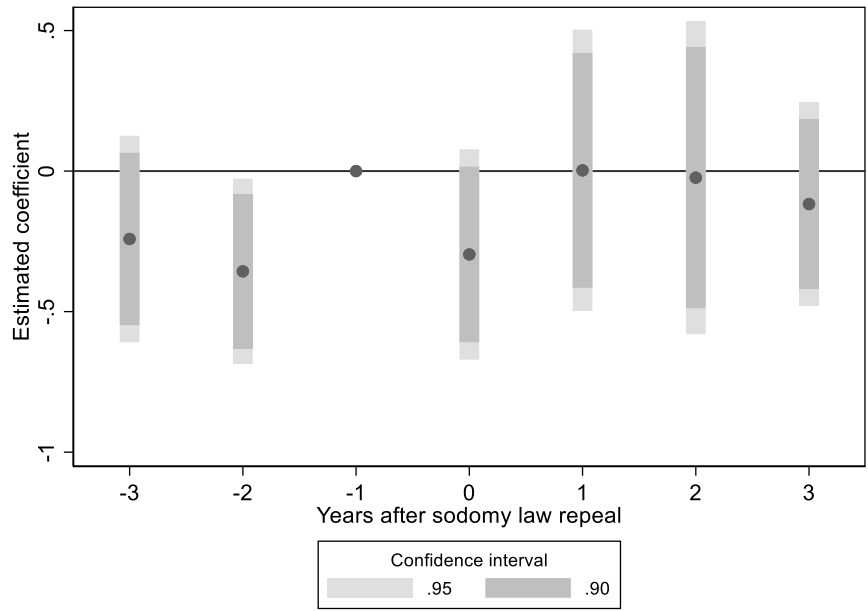

This figure analyzes the effect of sodomy law repeals on the arrest rate (in logarithm) for gambling. See also notes in Figure 1. Source: FBI 1995-2018. N=1,186.

**Figure C20: Effect of sodomy law repeals on arrests for burglary.**

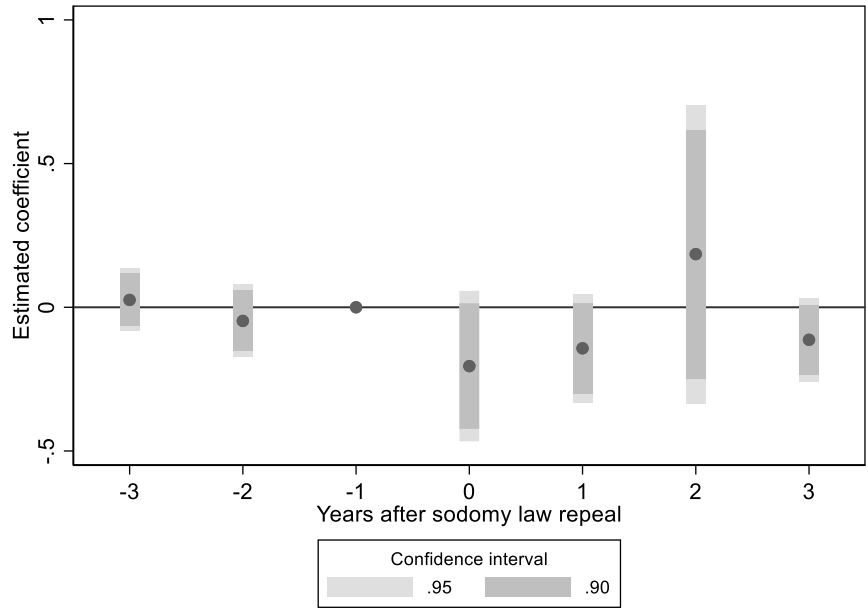

This figure analyzes the effect of sodomy law repeals on the arrest rate (in logarithm) for burglary. See also notes in Figure 1. Source: FBI 1995-2018. N=1,189.

**Figure C21: Effect of sodomy law repeals on arrests for arson.**

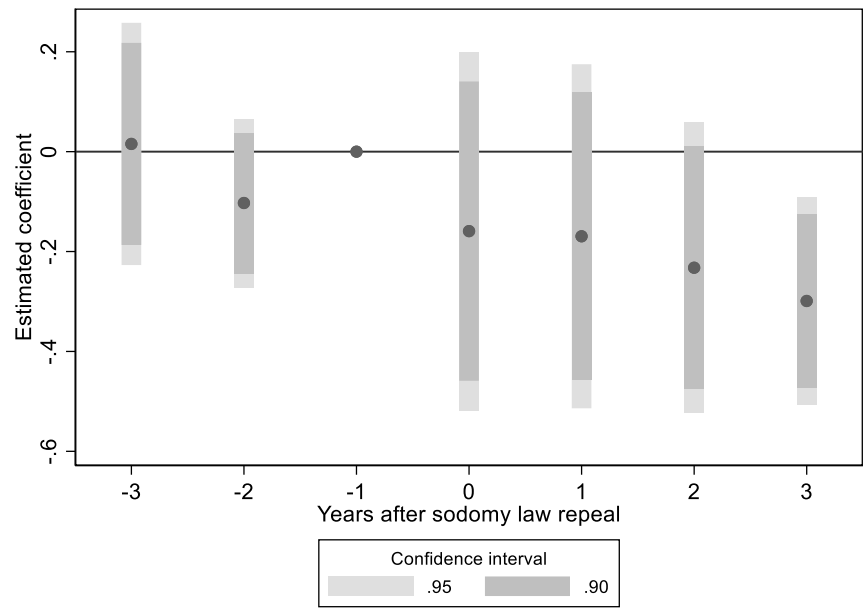

This figure analyzes the effect of sodomy law repeals on the arrest rate (in logarithm) for arson. See also notes in Figure 1. Source: FBI 1995-2018. N=1,189.

**Table C1: Number of agencies, descriptive statistics.**

|  | Mean | St. Dev. | Min | Max | Median |
|---|---|---|---|---|---|
| Alaska | 223.13 | 107.34 | 1 | 344 | 266.50 |
| Alabama | 29.08 | 3.90 | 19 | 35 | 30 |
| Arkansas | 83.75 | 7.08 | 67 | 98 | 84 |
| Arizona | 191.04 | 47.65 | 93 | 271 | 186.50 |
| California | 666.04 | 29.23 | 602 | 700 | 679.50 |
| Colorado | 173.67 | 25.73 | 125 | 208 | 180 |
| Connecticut | 94.88 | 7.84 | 74 | 105 | 96.50 |
| District of Columbia | 48.75 | 11.66 | 1 | 62 | 52 |
| Delaware | 1.50 | 0.52 | 1 | 2 | 1.50 |
| Florida | 596.95 | 52.81 | 475 | 678 | 595.50 |
| Georgia | 294.63 | 101.02 | 75 | 422 | 290.50 |
| Hawaii | 2.91 | 1.19 | 1 | 5 | 3 |
| Iowa | 100.75 | 10.84 | 71 | 117 | 102.50 |
| Idaho | 1.57 | 0.51 | 1 | 2 | 2 |
| Illinois | 158.08 | 36.55 | 102 | 243 | 155 |
| Indiana | 185.96 | 10.90 | 154 | 206 | 190 |
| Kansas | 222.47 | 16.83 | 183 | 248 | 225 |
| Kentucky | 160.75 | 168.33 | 3 | 419 | 29.50 |
| Louisiana | 130.25 | 24.40 | 86 | 170 | 130.50 |
| Massachusetts | 164.13 | 17.50 | 119 | 195 | 164 |
| Maryland | 144 | 7.03 | 129 | 154 | 144.50 |
| Maine | 293.71 | 30.45 | 239 | 342 | 301.50 |
| Michigan | 532.54 | 51.72 | 421 | 614 | 535.50 |
| Minnesota | 316.50 | 32.24 | 260 | 378 | 319 |
| Missouri | 76.67 | 16.58 | 41 | 101 | 77 |
| Mississippi | 317.46 | 127.43 | 132 | 580 | 354 |
| Montana | 76.80 | 22.51 | 34 | 100 | 89.50 |
| North Carolina | 198.46 | 43.21 | 54 | 237 | 212.50 |
| North Dakota | 32.83 | 9.93 | 3 | 51 | 32 |
| Nebraska | 128.50 | 37.43 | 40 | 184 | 132 |
| New Hampshire | 519.79 | 26.43 | 473 | 577 | 527.50 |
| New Jersey | 52.17 | 16.98 | 22 | 87 | 52.50 |
| New Mexico | 487.96 | 73.40 | 330 | 628 | 505.50 |
| Nevada | 338.46 | 73.61 | 204 | 463 | 337.50 |
| New York | 74.54 | 20.46 | 44 | 106 | 70 |
| Ohio | 354.63 | 74.29 | 231 | 461 | 358 |
| Oklahoma | 302.83 | 33.10 | 256 | 400 | 296 |
| Oregon | 146.50 | 22.52 | 101 | 194 | 144.50 |
| Pennsylvania | 905.83 | 329.88 | 1 | 1,383 | 885.50 |
| Rhode Island | 45.83 | 2.33 | 41 | 49 | 46.50 |
| South Carolina | 286.96 | 123.73 | 82 | 479 | 262.50 |
| South Dakota | 83.54 | 29.34 | 23 | 120 | 87 |
| Tennessee | 347.54 | 121.26 | 93 | 460 | 393.50 |
| Texas | 935.50 | 50.07 | 839 | 1,020 | 944 |
| Utah | 101.88 | 14.78 | 79 | 125 | 104.50 |
| Virginia | 59.59 | 16.55 | 18 | 78 | 65.50 |
| Vermont | 341.33 | 40.88 | 260 | 410 | 342.50 |
| Washington | 208.96 | 14.49 | 177 | 229 | 210 |
| Wisconsin | 230.13 | 63.44 | 126 | 347 | 248 |
| West Virginia | 326 | 92.85 | 3 | 427 | 342 |
| Wyoming | 57.71 | 7.40 | 31 | 65 | 61 |

This table displays descriptive statistics for the number of agencies in each state during the sample period. Source: FBI 1995-2018.

**Table C2: Effect of sodomy law repeals on arrests for sex offenses.**

|  | (1) | (2) | (3) |
|---|---|---|---|
| Time sodomy law repeal = -3 and earlier | -0.0292 | 0.00609 |  |
|  | (0.101) | (0.0898) |  |
| Time sodomy law repeal = -2 | -0.0724 |  | -0.0497 |
|  | (0.0923) |  | (0.0897) |
| Time sodomy law repeal = 0 | -0.258** | -0.224** | -0.236** |
|  | (0.110) | (0.100) | (0.117) |
| Time sodomy law repeal = +1 | -0.208** | -0.175** | -0.186 |
|  | (0.0987) | (0.0834) | (0.124) |
| Time sodomy law repeal = +2 | -0.0827 | -0.0481 | -0.0603 |
|  | (0.156) | (0.170) | (0.194) |
| Time sodomy law repeal = +3 and later | -0.207** | -0.173 | -0.186* |
|  | (0.101) | (0.108) | (0.111) |
| State fixed effects | ✓ | ✓ | ✓ |
| Year fixed effects | ✓ | ✓ | ✓ |
| LGBT policies | ✓ | ✓ | ✓ |
| State control | ✓ | ✓ | ✓ |
| Observations | 1,189 | 1,189 | 1,189 |
| F-test (p-value) |  | 0.946 | 0.582 |

See notes in Figure 1 and Table 1. *** p<0.01, ** p<0.05, * p<0.1

**Table C3: Effect of sodomy law repeals on arrests for prostitution.**

|                                       | (1)        | (2)        | (3)        |
|---------------------------------------|------------|------------|------------|
| Time sodomy law repeal = -3 and earlier | -0.0860    | -0.0193    |            |
|                                       | (0.196)    | (0.194)    |            |
| Time sodomy law repeal = -2           | -0.137     |            | -0.0701    |
|                                       | (0.212)    |            | (0.224)    |
| Time sodomy law repeal = 0            | -0.438*    | -0.374     | -0.374     |
|                                       | (0.261)    | (0.259)    | (0.232)    |
| Time sodomy law repeal = +1           | -0.625**   | -0.562**   | -0.561**   |
|                                       | (0.304)    | (0.247)    | (0.266)    |
| Time sodomy law repeal = +2           | -0.0217    | 0.0437     | 0.0444     |
|                                       | (0.381)    | (0.398)    | (0.375)    |
| Time sodomy law repeal = +3 and later | -0.641***  | -0.577**   | -0.578**   |
|                                       | (0.228)    | (0.232)    | (0.231)    |
| State fixed effects                   | ✓          | ✓          | ✓          |
| Year fixed effects                    | ✓          | ✓          | ✓          |
| LGBT policies                         | ✓          | ✓          | ✓          |
| State control                         | ✓          | ✓          | ✓          |
| Observations                          | 1,188      | 1,188      | 1,188      |
| F-test (p-value)                      |            | 0.921      | 0.756      |

See notes in Figure 1 and Table 1. *** p<0.01, ** p<0.05, * p<0.1

**Table C4: Effect of sodomy law repeals on arrests for disorderly conduct.**

|                                         | (1)        | (2)       | (3)        |
|-----------------------------------------|------------|-----------|------------|
| Time sodomy law repeal = -3 and earlier | 0.0444     | 0.0814    |            |
|                                         | (0.0879)   | (0.0823)  |            |
| Time sodomy law repeal = -2             | -0.0761    |           | -0.111     |
|                                         | (0.0682)   |           | (0.0784)   |
| Time sodomy law repeal = 0              | -0.295***  | -0.259**  | -0.328**   |
|                                         | (0.101)    | (0.101)   | (0.123)    |
| Time sodomy law repeal = +1             | -0.298**   | -0.263**  | -0.332**   |
|                                         | (0.113)    | (0.108)   | (0.134)    |
| Time sodomy law repeal = +2             | -0.121     | -0.0848   | -0.155     |
|                                         | (0.152)    | (0.160)   | (0.178)    |
| Time sodomy law repeal = +3 and later   | -0.273***  | -0.238**  | -0.306***  |
|                                         | (0.0973)   | (0.0989)  | (0.108)    |
| State fixed effects                     | ✓          | ✓         | ✓          |
| Year fixed effects                      | ✓          | ✓         | ✓          |
| LGBT policies                           | ✓          | ✓         | ✓          |
| State control                           | ✓          | ✓         | ✓          |
| Observations                            | 1,179      | 1,179     | 1,179      |
| F-test (p-value)                        |            | 0.327     | 0.164      |

See notes in Figure 1 and Table 1. *** $p<0.01$, ** $p<0.05$, * $p<0.1$

**Table C5: Effect of sodomy law repeals on arrests for driving under the influence.**

|  | (1) | (2) | (3) |
|---|---|---|---|
| Time sodomy law repeal = -3 and earlier | 0.0614 | 0.0533 | |
|  | (0.0799) | (0.0710) | |
| Time sodomy law repeal = -2 | 0.0166 | | -0.0311 |
|  | (0.0463) | | (0.0554) |
| Time sodomy law repeal = 0 | -0.177** | -0.185** | -0.223** |
|  | (0.0847) | (0.0886) | (0.109) |
| Time sodomy law repeal = +1 | -0.260*** | -0.268*** | -0.306** |
|  | (0.0929) | (0.0990) | (0.123) |
| Time sodomy law repeal = +2 | -0.115 | -0.123 | -0.162 |
|  | (0.204) | (0.203) | (0.210) |
| Time sodomy law repeal = +3 and later | -0.299*** | -0.307*** | -0.345*** |
|  | (0.0858) | (0.0837) | (0.0889) |
| State fixed effects | ✓ | ✓ | ✓ |
| Year fixed effects | ✓ | ✓ | ✓ |
| LGBT policies | ✓ | ✓ | ✓ |
| State control | ✓ | ✓ | ✓ |
| Observations | 1,188 | 1,188 | 1,188 |
| F-test (p-value) |  | 0.456 | 0.577 |

See notes in Figure 1 and Table 1. *** $p<0.01$, ** $p<0.05$, * $p<0.1$